\documentclass{aa}
\usepackage{graphicx}
\usepackage{txfonts}
\usepackage{xcolor}
\usepackage{caption}
\usepackage{subcaption}
\usepackage{textcomp}
\usepackage[rightcaption]{sidecap}
\usepackage[mathlines]{linenoaa}
\usepackage{placeins}  
\usepackage{media9}
\usepackage{hyperref}

\usepackage{multirow}

\newcommand{\CellWithForceBreak}[2][c]{
\begin{tabular}[#1]{@{}c@{}}#2\end{tabular}}

\begin{document} 

    \title{Magnetic flux systems involved in the May 2024 solar energetic events from AR~13664 inferred through coronal dimmings}
    \titlerunning{Magnetic flux systems involved in the May 2024 solar energetic events} 
        
    \author{Amaia Razquin
          \inst{1}
          \and
          Karin Dissauer\inst{1, 3}
          \and
          Astrid M. Veronig\inst{1, 2}
          \and 
          Graham Barnes\inst{3}
    }

    \institute{University of Graz, Institute of Physics, Universitätsplatz 5, 8010 Graz, Austria \\
        \email{amaia.razquin-lizarraga@uni-graz.at}
        \and
             NorthWest Research Associates, 3380 Mitchell Lane, Boulder, CO 80301, USA
        \and 
            University of Graz, Kanzelh\"ohe Observatory for Solar and Environmental Research, Kanzelh\"ohe 19, 9521 Treffen, Austria
    }
        
    \date{Received January, 2026; accepted March, 2026}
        
    \abstract 
    {Coronal dimmings are transient depletions of coronal plasma observed in extreme ultraviolet and soft X-ray wavelengths interpreted as low-corona signatures of coronal mass ejections (CMEs). Their evolution is closely linked to CME dynamics, flare reconnection, and the large-scale reconfiguration of the coronal magnetic field. During May 2024, the active region (AR) 13664 produced 66 $\ge$ M-class flares alongside a sequence of fast CMEs that caused the largest geomagnetic storm since 2003. AR~13664 was also among the largest active regions and contained one of the largest amounts of magnetic flux ever recorded. It provided an exceptional opportunity to study the magnetic coupling between dimmings, flares, and CMEs within a single highly active AR.
    }
    {We investigate the morphology, magnetic properties, and temporal evolution of coronal dimmings produced by AR 13664. We expand on a previous paper where we identified all coronal dimmings from AR~13664 and performed a statistical analysis of their characteristic parameters in relation to the associated flares and CMEs to determine how the spatial development of the dimmings relates to flare ribbon locations and to the magnetic field configuration of the AR. We aim to identify the magnetic flux systems involved in the eruptions and assess how the observed dimming evolution reflects large-scale coronal restructuring and CME initiation.
    }  
    {We analysed 16 on-disc coronal dimmings from AR 13664 detected in SDO/AIA 211~\AA~observations between May 1 and 14, 2024. We extracted coronal dimmings using logarithmic base-ratio thresholding, and their magnetic properties were derived from SDO/HMI line-of-sight magnetograms. We detected flare ribbons in AIA 1600~\AA~data using an adaptive thresholding technique and computed their reconnection fluxes from radial magnetic field maps. To explore the magnetic environment, we used high-resolution potential field source surface (PFSS) and non-linear force-free (NLFF) extrapolations of the coronal magnetic field and traced the magnetic flux systems connecting dimming regions, flare ribbons, and coronal holes.
    }
    {We found strong correlations between flare ribbon and dimming parameters. The magnetic dimming area and flare ribbon area correlate with $c=0.65\pm0.13$, and the unsigned dimming and reconnection fluxes correlate with $c=0.60\pm0.18$, consistent with earlier statistical studies. The morphology of the dimmings changed systematically over the evolution of the AR, with southwards expanding dimmings occurring before May 9 and northwards expanding ones thereafter. This transition coincides with a change in the flare ribbon locations. AR~13664 contains two long, strong, and almost horizontal, i.e. east-west, polarity inversion lines (PILs), and the flare ribbon locations shift from the southern to the northern PIL. Together with the dimming location, these changes imply the presence of two distinct magnetic domains. 
    The PFSS extrapolations showed that southward (northward) dimmings are mainly strapping flux dimmings with magnetic field lines vaulting above the southern (northern) PIL. The final extent of the dimmings was then given by the exterior flux involved in the eruption via stretching and reconnection. In one event, we found an extended quiet Sun dimming potentially triggered by field line opening due to the passage of an extreme ultraviolet wave. 
    }
    {}
        
    \authorrunning{A. Razquin et al.}
    \maketitle 
\nolinenumbers

\section{Introduction}
Coronal dimmings are transient decreases in extreme ultraviolet (EUV) and soft X-ray (SXR) emission from the solar corona interpreted as signatures of rapid plasma evacuation and density depletion during the early evolution of coronal mass ejections (CMEs; \citealt{hudson1996long, sterling1997yohkoh,  thompson1998soho, thomson2000coronal}). This interpretation is supported by evidence from multiwavelength imaging studies \citep{Zarro1999soho}, spectroscopic analyses \citep{harra2001material, Jin2009coronal, Veronig2019}, and differential emission measure diagnostics \citep{vanninathan2018plasma}. For a recent review on the properties of coronal dimmings and their relation to CMEs, we refer to \citet{Veronig2025}.

Statistical analyses have established correlations between on-disc coronal dimming parameters and CME properties -- including mass, velocity, and acceleration -- as well as with the characteristics of associated flares \citep{reinard2009relationship, aschwanden2016global, krista2017statistical, Dissauer2018b, Dissauer2019, arazquin2025coronal}. Off-limb dimmings have likewise been statistically associated with CMEs \citep{bewsher2008relationship, aschwanden2017global, chikunova2020coronal}. Dimmings have also been identified in spatially integrated observations, so-called Sun-as-a-star observations \citep{mason2014mechanisms, mason2016relationship}, and successfully applied in stellar CME detection \citep{ Veronig2021indications, loyd2022constraining}. 

Recent studies have shown that dimming expansion provides insights into the CME expansion direction \citep{Chikunova2023, Podladchikova2024, jain2024coronal} and recovery processes \citep{kahler2001origin, attrill2008recovery, vanninathan2018plasma, ronca2024recovery}. As such, the dimming inferred estimation of CME direction (DIRECD) method was developed to use dimming expansion to estimate the CME propagation direction \citep{jain2024estimating, jain2025validating}.

Traditionally, coronal dimmings have been classified into two groups based on their visual appearance on the solar disc: Core dimmings are localised dimmings, often found in magnetically conjugated pairs, with a pronounced decrease in brightness, while secondary dimmings are extended and shallow. Core dimmings are understood to mark the footpoints in different polarities of the erupting flux rope \citep{thompson1998soho, attrill2006using}, whereas secondary dimmings show the density depletion that results from the stretching and partial reconnection of overlying magnetic fields \citep{mandrini2007cme, Dissauer2018b}. To increase the diagnostic potential of dimmings in eruptive processes, flare ribbons, which trace the footpoints of newly reconnected field lines in the photosphere, can be used \citep{fletcher2011observational, Kazachenko2017database}. \citet{Veronig2025} introduced a characterisation of dimmings derived from the spatiotemporal development of dimming regions in connection with flare ribbon evolution. This new categorisation enables the connection of the observed plasma depletion to specific magnetic flux systems involved in the eruption and allows for the assessment of the global reconfiguration of the corona in association with the eruption.

In May 2024, the NOAA (National Oceanic and Atmospheric Administration) active region (AR) 13664 generated the strongest geomagnetic storm since 2003 \citep{kwak2024observational, Weiler2024first, hayakawa2024solar}. The region was visible from May 1, and it rapidly grew to a size comparable to the AR that generated the Carrington event \citep{carrington1859description} until it rotated over the western limb on May 14. It evolved from a bipolar sunspot group into a Fkc McIntosh-type of the magnetic class $\beta\gamma\sigma$ starting on May 6 \citep{hayakawa2024solar}. After May 7, the magnetic flux and free energy of AR~13664 strongly increased, reaching magnetic flux values of up to $1.9\times10^{23}$~Mx \citep{Jarolim2024magnetic, hayakawa2024solar}. Between May 8 and 14, it produced 14 major CMEs, ten of them halo CMEs, leading to a geomagnetic storm that peaked at a Dst index of $-412$~nT on May 11.
\citet{arazquin2025coronal} performed coronal dimming detection on 1 C-, 54~M- and 12~X-class flares from AR~13664 between May 1 and May 14, and they found 16 on-disc and six off-limb coronal dimmings. In addition, they associated the flares with 23~CMEs, from which ten were Earth-directed CMEs. Their study revealed exceptionally large dimming magnetic fluxes ($>2.75\times10^{21}$~Mx) and areas ($>1.16\times10^{10}$~km$^2$) as well as strong correlations between dimming properties, flare parameters, and CME speeds.

Building on these results, we focus on the morphology and temporal evolution of the coronal dimmings from AR~13664 in relation to flare ribbon locations and the magnetic field configuration of the AR in this study. We apply a potential field source surface (PFSS) model to extract the large-scale coronal structures overlying AR 13664 and investigate their role in shaping the dimming signatures. We use non-linear force-free (NLFF) extrapolations to understand the contribution of the flux rope to the dimming regions. By relating the dimming evolution to the flare ribbon development and embedding both diagnostics into the classification scheme of \citet{Veronig2025}, we aim to elucidate the magnetic flux systems involved in the eruptions and to advance the interpretation of dimmings as diagnostics of CME initiation and coronal magnetic field reconfiguration.

\section{Dataset and data reduction} \label{sec:data}
We study the on-disc coronal dimmings from AR~13664 between May 1 and May 14 detected by \citet{arazquin2025coronal}. This comprises a dataset of 16 coronal dimmings from which six are associated with X-class flares, nine with M-class flares, and one with a C-class flare. Thirteen of these coronal dimmings are accompanied by a CME, seven of which were halo CMEs. A detailed description of these events and how they were detected can be found in \citet{arazquin2025coronal}. Table~\ref{table:dimming_properties} gives an overview of the timing, location, and strength of these events, along with the maximum speeds of the associated CMEs derived from height–time profiles in the CDAW SOHO/LASCO catalogue\footnote{\url{https://cdaw.gsfc.nasa.gov/CME_list/}} \citep{arazquin2025coronal}, and an indication of whether they were halo CMEs.

To analyse coronal dimmings we used full-cadence (12~s) 211~\AA~images of the Solar Dynamics Observatory (SDO; \citealt{pesnell2012sdo})/Atmospheric Imaging Assembly (AIA; \citealt{Lemen2012aia}), and 720~s line-of-sight (LOS) magnetograms from the SDO/Helioseismic and Magnetic Imager (HMI; \citealt{Scherrer2012HMI, Schou2012design}). The images were re-binned to $2048\times2048$ pixels under flux conservation conditions, resulting in an effective spatial resolution of 1.2~arcsecs per pixel, and processed with standard Solarsoft IDL software (\texttt{aia\_prep.pro} and \texttt{hmi\_prep.pro}). Furthermore, \texttt{coreg\_map.pro} is used to co-register the AIA and HMI data for magnetic flux calculations of the dimming regions. For each event, we differentially rotated the images to a reference frame 15~min before the associated flare onset in order to correct for solar differential rotation. We analysed the dimmings over a 2~hr~15~min period, starting 15~min before the flare. We only used AIA images with exposures of 1.8–3.0~s. For the magnetic properties of the dimmings we used a single magnetogram taken 15~min before the associated flare onset. We studied coronal dimmings within a $1000\times 1000''$ subfield around the AR for the extraction of the characteristic parameters, and a second run was done in full-frame images to understand the total extent of the coronal dimmings. 

To identify the flare ribbons we used AIA 1600~\AA~images with 24~s cadence. Similarly to the 211~\AA~images, we rebinned the 1600~\AA~images to $2048\times2048$ pixels and processed them with \texttt{aia\_prep.pro}. We performed a visual inspection for saturated pixels and excluded any images in which they were present. We detected the flare ribbons from the start time to the end time of the flare, except for event no.~3, which is a M2.6 flare on May 7 00:41~UT, where we initialised the detection at 00:00~UT. This adjustment was necessary because the GOES 1-8~\AA~flux remains above $10^{-5}$~Wm$^{-2}$ from May 6 22:00~UT onward, due to activity originating from AR~13663 in the northern hemisphere. As a result, distinguishing the precise onset of the flare from AR~13664 was not feasible, and consequently we allowed the detection to start earlier, namely, co-temporally with the brightenings observed in 211~\AA~images. To extract the ribbon reconnection flux, we used HMI vector magnetograms (\texttt{hmi.B\_720s}). We computed the radial magnetic field $B_r$ from the vector magnetograms using the IDL procedure \texttt{hmi\_b2ptr}\footnote{\url{http://jsoc.stanford.edu/data/hmi/ ccmc/}} following \citet{sun2013coordinate} and rebinned the $B_r$ maps to $2048\times2048$ pixels. Both the radial magnetic field maps and the 1600~\AA~images were differentially rotated to match the detection frame of the dimming regions. We note that we used different magnetograms for the dimmings ($B_\text{LOS}$) and for the flare ribbons ($B_r$) to be consistent with the methods employed in \citet{Dissauer2018b}, as the data from that study was directly compared to the one in this study in Sect.~\ref{sec:results:cor}.

\section{Methods} \label{sec:methods}
\subsection{Coronal dimming detection}
\begin{figure}[t] 
\centering
    \resizebox{\hsize}{!}{\includegraphics{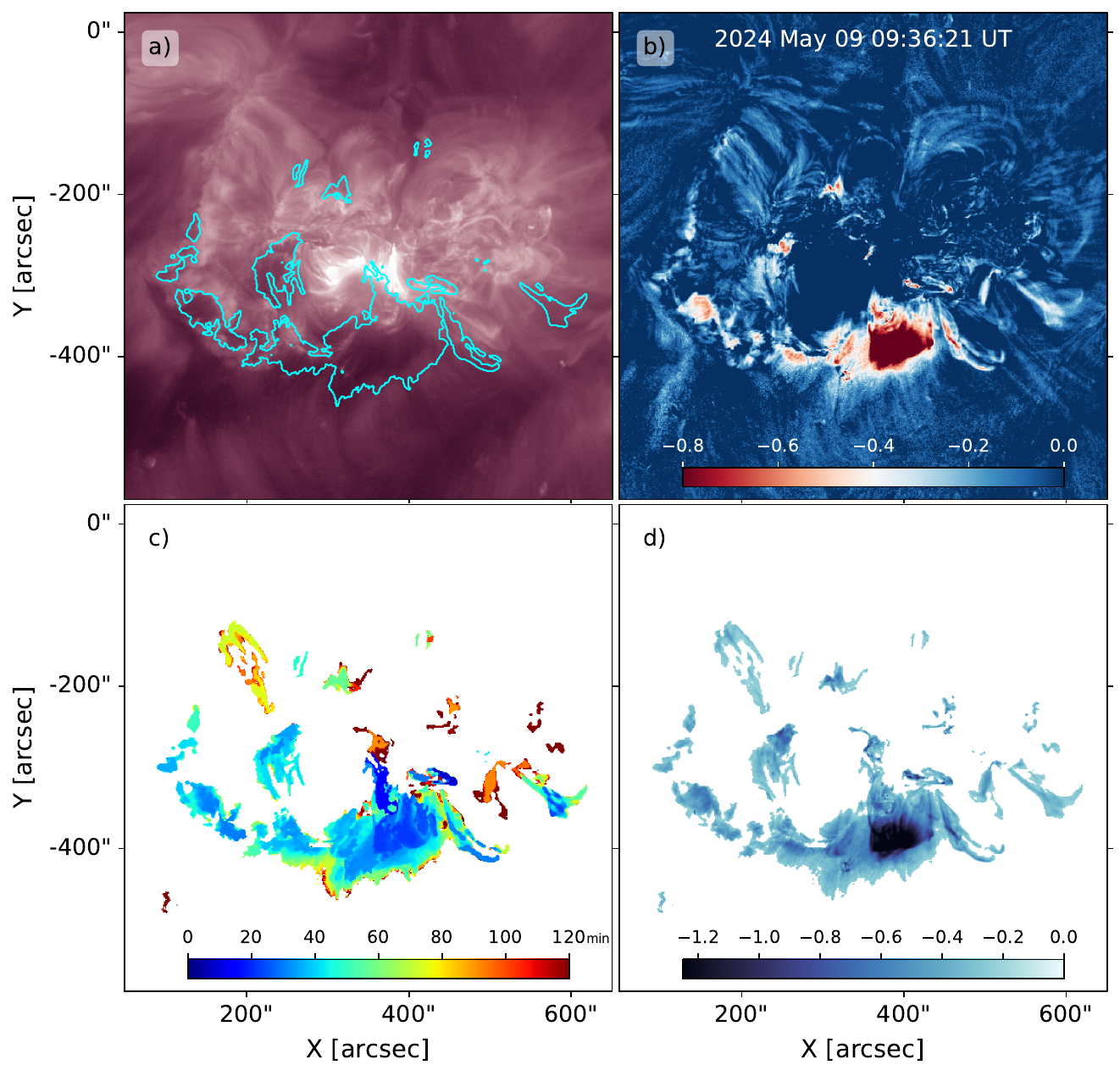}}
        \caption{Overview of the evolution of the X2.2 flare and associated dimming on May 9, 2024, at 08:45~UT. (a) SDO/AIA 211~\AA~ direct images of AR~13664 at the flare’s end time, with the dimming region detected up to that point outlined in cyan contours. (b) Corresponding logarithmic base-ratio image. (c) Timing map indicating the first detection time in minutes of a dimming pixel. (d) Minimum intensity map from logarithmic base-ratio data. The associated movie is available online.}
        \label{fig:methods:dimming}
\end{figure}

We extract coronal dimmings using the method developed by \citet{Dissauer2018a} where a thresholding technique is applied on AIA 211~\AA~images. In this method, a pixel is considered a dimming pixel if its logarithmic base-ratio (log$_\text{10}$) intensity decreases below $-0.19$. In this study, we construct the logarithmic base-ratio images using as a base image the average of three consecutive images 15~min before the flare onset. Pixels that fulfil the condition are cumulatively added to a dimming pixel mask. Figure~\ref{fig:methods:dimming} shows the overview of the coronal dimming detection for the X2.2 flare on May 9 at 08:45~UT to illustrate the dimming detection method. Panel a shows the dimming pixel mask in cyan contours on top of the 211~\AA~image at the end of the flare (09:36~UT), while panel b shows the corresponding logarithmic base-ratio image at the same time step. In Fig.~\ref{fig:methods:dimming}b the white and red regions mark areas where the brightness has strongly decreased; that is, the dimming regions. 

In addition, we create a timing map in which we record the first time a pixel is detected as a dimming pixel which shows the spatial expansion of the dimming in time. Figure~\ref{fig:methods:dimming}c shows an example of a timing map with the dark blue regions indicating where the dimming occurs first. Similarly, we create minimum intensity maps that show the minimum intensity over the detection time of the dimming pixels. A minimum intensity map is shown in Fig.~\ref{fig:methods:dimming}d, where darker pixels are linked to larger mass depletions. The temporal evolution of this figure is available as an online movie.
     
In order to compare the coronal dimming properties to the flare ribbon properties, we extracted the magnetic dimming area $A_\phi$ which is defined as the final cumulative area of all dimming pixels that are in regions where the magnetic field $B$ in the HMI LOS magnetogram is above the noise level, $|B|> 10$~G. We assumed that the magnetic field is radial and applied the `$\mu$-correction', thereby $B=B_{\text{LOS}}/cos(\theta)$ where $\theta$ is the observing angle. From this region we also extracted the dimming unsigned magnetic flux $\phi$. For more details on this method, we refer to \citet{Dissauer2018a}. \citet{arazquin2025coronal} gives an in-depth description of the coronal dimming detection for AR~13664 and the extraction of all characteristic parameters.

\subsection{Flare ribbon detection}
We identified the flare ribbons using a thresholding technique based on the method developed by \citet{Kazachenko2017database}. We applied a threshold to AIA 1600~\AA~images and identify all pixels with an intensity above a cutoff value $I_c$. The cutoff intensity $I_c$ is defined as the median image intensity times $c$, where $c\in [6,8,10]$. We cumulatively added every newly detected flare ribbon pixel to a mask from which we extract the area of the flare ribbons $A_c^{rbn}$. We used the central cutoff value $I_8$ to define the area of the flare ribbons $A_{rbn}=A_8^{rbn}$ and computed the uncertainty from the standard deviation of the areas extracted with all cutoff values together. 

We calculated the ribbon reconnection flux, i.e. the magnetic flux swept by the flare ribbons, using the radial magnetic field from the HMI vector magnetograms. We calculated the ribbon reconnection flux $\phi_c^{rbn}$ as the flux within the flare ribbon area $A_c^{rbn}$ where the HMI radial magnetogram surpasses the noise level $|B_r|>100$~G. 
Similarly to the ribbon area, we define the ribbon unsigned magnetic flux from the central cutoff value ($I_8$) as $\phi_{rbn}=\phi_8^{rbn}$ and get the uncertainties from the standard deviation of the ribbon fluxes extracted with all ribbon areas. 

\subsection{Magnetic field extrapolations} 
To investigate the magnetic field structures within the dimming regions, we employed two different magnetic field extrapolations. We used a PFSS model (\citealt{altchuler1969magnetic, schatten1969magnetic}) to capture the large-scale magnetic domains, whereas we used NLFF extrapolations to resolve the non-potential magnetic structures within the AR.

We used PFSS extrapolations to explore the magnetic field configuration in the extended dimming regions. To resolve small spatial scales, we followed the procedure presented by \citet{dissauer2025uniqueness} and \citet{barnes2026}, where the calculations are performed using the SHTools package \citep{wieczorek2018shtools}. This package accurately computes the spherical harmonics needed for the PFSS model up to a maximum degree of roughly 2800. For the boundary condition, we used a full-disc map of the radial magnetic field from the HMI vector magnetogram ($B_{r}$), assuming an anti-symmetry condition on the far side of the Sun.

We displayed the results using the ParaView\footnote{\url{https://www.paraview.org}} open-source visualisation software. For each flare, we overlaid LOS magnetograms recorded 15~min before flare onset with shaded contours of the coronal dimmings and flare ribbons. We used the coronal dimmings detected on the full-disc to understand the full extent of the magnetic flux systems involved. We placed seed sources for stream tracing within the coronal dimming contours to visualise the magnetic field lines connecting to the dimming regions of interest. We limit the seed sources to be within the dimming regions detected during the impulsive phase of the events. We note that ParaView does not natively support spherical grids, consequently, we converted the PFSS extrapolation to cartesian coordinates. This transformation can introduce uncertainties, particularly in regions near the limb.

The PFSS models are current-free and thus unable to include structures such as flux ropes in their extrapolations. To identify potential core dimming regions, i.e. footpoints of the erupting flux ropes, we used the NLFF extrapolations of AR~13664 made by \citet{Jarolim2024magnetic}. They employed the method from \citet{jarolim2023probing}, based on physics-informed neural networks, to extract the evolution of the magnetic field configuration of AR~13664 between May~5 and May~11. We used a selection of their publicly available extrapolations\footnote{\url{https://app.globus.org/file-manager?origin_id=4263de78-cfdb-401e-a62b-dae3b935530a&origin_path=\%2F&two_pane=false}} and visualised them using ParaView. We placed seed sources for stream tracing within the flare ribbon region to visualise the flux ropes.

\section{Results} \label{sec:results}
Table~\ref{table:dimming_properties} summarises all the on-disc coronal dimmings originating from AR~13664 analysed in this study and identified by \citet{arazquin2025coronal}. In total, these are 16 dimming events that occurred between May 5 and 11, 2024. Six of the dimmings are associated with X-class flares, nine with M-class flares and one with a C-class flare. Except for three M-class flares, all studied flares during this time span have been associated with a CME, seven of which were halo CMEs. Characteristic magnetic parameters of the coronal dimmings and flare ribbons derived as described in Sect.~\ref{sec:methods} are summarised in Table~\ref{table:dimming_properties}.
    
\subsection{Correlation of flare ribbon and coronal dimming characteristic parameters} \label{sec:results:cor}

\begin{table*}
    \caption{\label{table:dimming_properties}Coronal dimming and flare ribbon properties of the on-disc coronal dimmings from AR~13664.}   
    \centering
    \begin{tabular}{ccccccccccc}         
    \hline\hline          
    N & Day & Start & \CellWithForceBreak{Flare \\ Location}& \CellWithForceBreak{F$_P$ \\ W m$^{-2}$} &  \CellWithForceBreak{$A_\phi$ \\ $10^{10} \text{km}^2$} &  \CellWithForceBreak{$\phi$ \\ $10^{21} \text{Mx}$} &  \CellWithForceBreak{$A_{rbn}$ \\ $10^{8} \text{km}^2$} &  \CellWithForceBreak{$\phi_{rbn}$ \\ $10^{21} \text{Mx}$} & \CellWithForceBreak{$v_\text{CME}$ \\ (km s$^{-1}$)} & \CellWithForceBreak{Dimming \\ Expansion} \\
    \hline
    $1$ & 5 & 14:33 & S20 E16 &  7.54E-06  & $ 0.78\pm 0.11 $ & $ 2.81\pm 0.48 $ & $ 0.62\pm 0.23 $ & $ 0.47\pm 0.12 $ & $623$ & south, north  \\
    $2$ & 5 & 18:34 & S17 E21 &  1.07E-05  & $ 0.64\pm 0.09 $ & $ 2.77\pm 0.29 $ & $ 0.88\pm 0.35 $ & $ 0.47\pm 0.11 $ & ... & south  \\
    $3$ & 7 & 00:41 & S18 E6 &  2.67E-05  & $ 0.75\pm 0.10 $ & $ 4.15\pm 0.51 $ & $ 1.23\pm 1.43 $ & $ 0.61\pm 0.25 $ & $484$ & south-east  \\
    $4$ & 7 & 11:40 & S19 E0 &  2.49E-05  & $ 0.54\pm 0.09 $ & $ 4.25\pm 0.98 $ & $ 2.89\pm 0.94 $ & $ 1.94\pm 0.40 $ & $445$ & south  \\
    $5$ & 8 & 03:19 & S18 W6 &  1.91E-05  & $ 0.82\pm 0.14 $ & $ 4.49\pm 0.67 $ & $ 1.53\pm 0.92 $ & $ 1.08\pm 0.39 $ & $812$ & south  \\
    $6^\dagger$ & 8 & 04:37 & S19 W7 &  1.04E-04  & $ 1.46\pm 0.17 $ & $ 7.37\pm 0.86 $ & $ 8.95\pm 2.39 $ & $ 6.78\pm 1.27 $ & $1103$ & south  \\
    $7^\dagger$ & 8 & 11:13 & S17 W8 &  8.60E-05  & $ 0.88\pm 0.11 $ & $ 4.91\pm 0.73 $ & $ 3.32\pm 1.30 $ & $ 1.91\pm 0.62 $ & $1321$ & south-east  \\
    $8$ & 8 & 19:15 & S19 W19 &  2.08E-05  & $ 0.76\pm 0.12 $ & $ 10.14\pm 1.47 $ & $ 1.15\pm 0.73 $ & $ 0.78\pm 0.42 $ & ... & south  \\
    $9^\dagger$ & 8 & 21:12 & S19 W21 &  1.00E-04  & $ 1.28\pm 0.20 $ & $ 13.11\pm 1.53 $ & $ 6.77\pm 4.23 $ & $ 4.00\pm 2.70 $ & $1304$ & south, north  \\
    $10$ & 9 & 06:03 & S19 W23 &  2.35E-05  & $ 0.34\pm 0.05 $ & $ 2.84\pm 0.49 $ & $ 2.94\pm 1.08 $ & $ 2.23\pm 0.46 $ & ... & north, south  \\
    $11^\dagger$ & 9 & 08:45 & S19 W22 &  2.24E-04  & $ 1.03\pm 0.15 $ & $ 8.55\pm 1.02 $ & $ 18.47\pm 4.50 $ & $ 12.59\pm 2.51 $ & $1963$ & south  \\
    $12$& 9 & 11:50 & S16 W33 &  3.15E-05  & $ 0.56\pm 0.10 $ & $ 3.31\pm 0.95 $ & $ 8.74\pm 2.07 $ & $ 4.10\pm 0.89 $ & $1036$ & south  \\
    $13^\dagger$ & 9 & 17:23 & S16 W27 &  1.12E-04  & $ 1.65\pm 0.22 $ & $ 8.27\pm 2.13 $ & $ 10.20\pm 2.21 $ & $ 6.41\pm 1.15 $ & $1454$ & north, south  \\
    $14^\dagger$ & 10 & 06:27 & S16 W33 &  3.95E-04  & $ 2.07\pm 0.25 $ & $ 6.83\pm 1.48 $ & $ 15.55\pm 3.21 $ & $ 11.11\pm 1.72 $ & $1249$ & north, south  \\
    $15^\dagger$ & 11 & 01:10 & S16 W43 &  5.88E-04  & $ 3.08\pm 0.33 $ & $ 8.88\pm 1.17 $ & $ 24.24\pm 3.96 $ & $ 15.32\pm 2.10 $ & $1982$ & north, south  \\
    $16$ & 11 & 14:46 & S14 W48 &  8.93E-05  & $ 1.21\pm 0.17 $ & $ 7.65\pm 2.46 $ & $ 15.18\pm 4.34 $ & $ 7.77\pm 1.63 $ & $1058$ & north  \\
    \hline 
    \end{tabular}
    \tablefoot{We list the day, start time, location, and peak SXR flux of the associated flare derived from the GOES flare catalogue. We list the magnetic dimming area ($A_\phi$), the total unsigned magnetic flux of the dimming ($\phi$), the flare ribbon area ($A_{rbn}$), and the flare ribbon flux ($\phi_{rbn}$). In the `Dimming Expansion' column we list the dimming expansion direction as described in Sect.~\ref{sec:location}. We list the maximum velocity of the associated CME ($v_\text{CME}$) as given in \citet{arazquin2025coronal}. Events marked with a ($^\dagger$) are associated with a halo CME.}
\end{table*}

\begin{figure}
    \begin{subfigure}{\columnwidth}
        \centering
        \resizebox{\hsize}{!}{\includegraphics{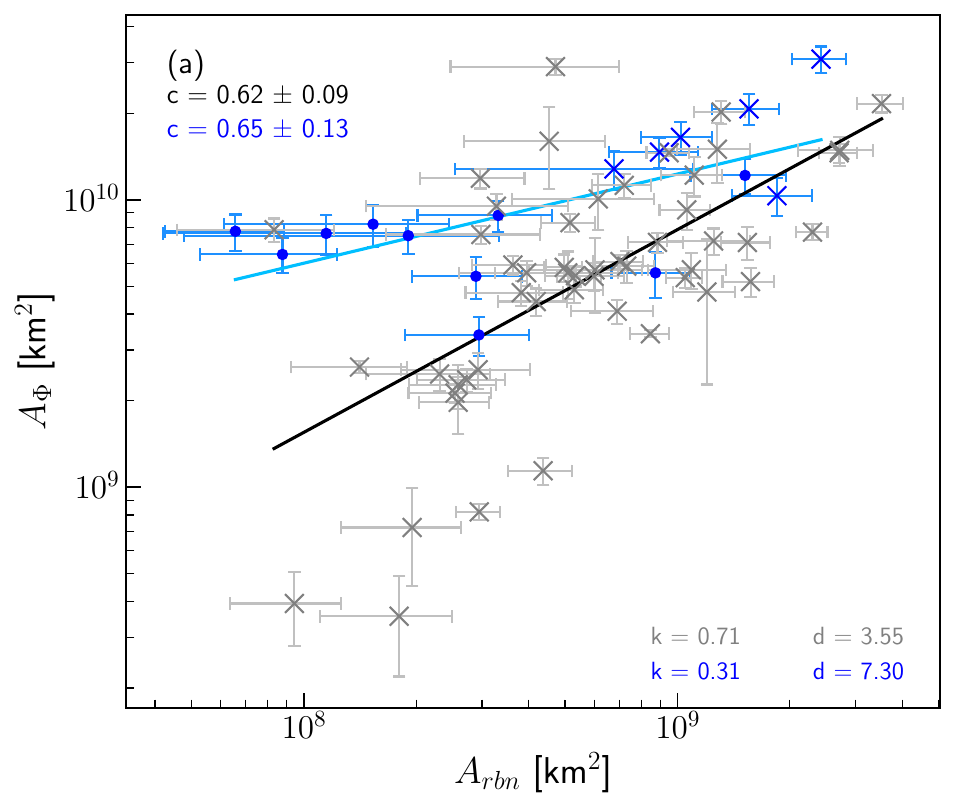}}
    \end{subfigure}
    \hfill
    \begin{subfigure}{\columnwidth}
        \centering
        \resizebox{\hsize}{!}{\includegraphics{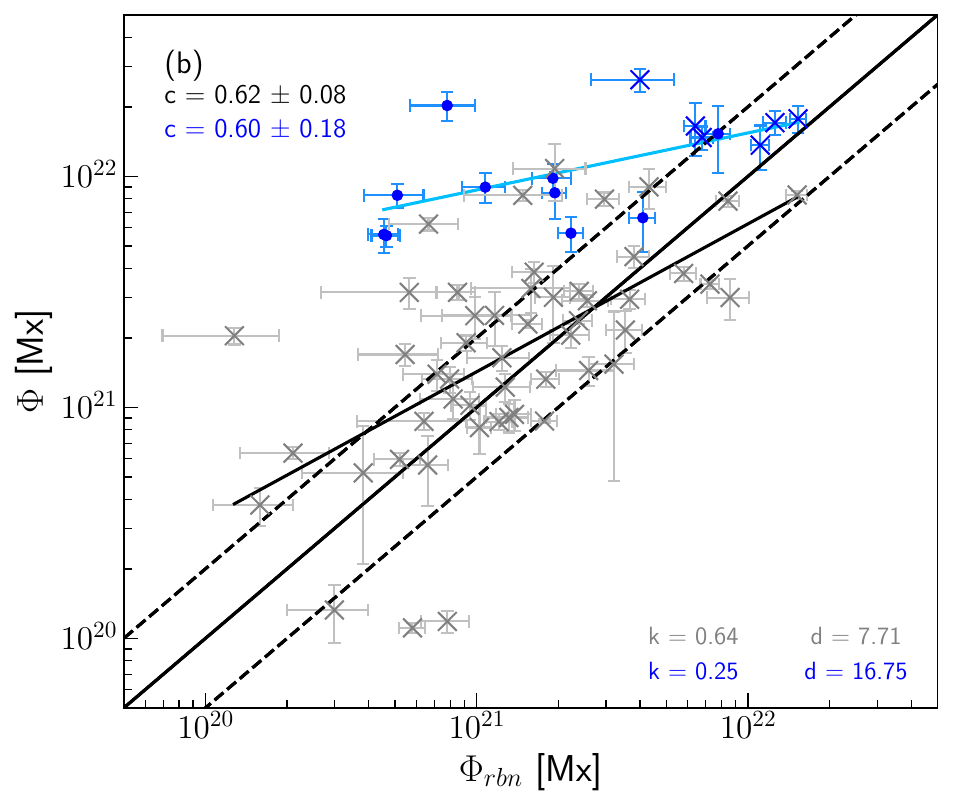}}
    \end{subfigure}
    \caption{(a) Comparison of the magnetic dimming area ($A_\phi$) against the flare ribbon area ($A_{rbn}$). (b) Total unsigned magnetic dimming flux ($\phi$) compared to the reconnection magnetic flux ($\phi_{rbn}$). Blue markers represent dimmings from the May 2024 events (crosses mark X-class flares and dots M-class flares), while grey crosses correspond to dimming events from \citet{Dissauer2018b}. The black (blue) regression lines were exclusively fitted to the grey (blue) data points.}
    \label{fig:cor}
\end{figure}

To investigate the connection between flare ribbons and coronal dimmings, we compared flare ribbon reconnection fluxes and areas with the corresponding magnetic fluxes and areas of coronal dimmings. Previous studies \citep{qiu2007magnetic, Dissauer2018b} have shown strong correlation ($r\approx0.6$) between the magnetic flux inferred from coronal dimmings and the reconnection flux derived from flare ribbons. This empirical relationship supports the hypothesis by  \citet{lin2004role} that, during a CME eruption, equal amounts of magnetic flux leave both ends of the reconnecting current sheet. In this scenario, the flux associated with coronal dimmings represents the overlying fields stretched and closed down by magnetic reconnection, which subsequently contributes poloidal flux to the erupting flux rope, while the reconnection flux traced by flare ribbons reflects the formation of newly closed loops beneath the current sheet. 
    
In order to test whether this hypothesis holds true on a single evolving AR, Fig.~\ref{fig:cor} shows the relevant flare ribbon and dimming parameters for the May 2024 events dimmings. We compare these parameters to those found for the general dimming population studied by \citet{Dissauer2018b}. To ensure consistency with that study and enable a direct comparison, we used different magnetograms to compute the properties of the flare ribbons and the coronal dimmings (see Sect.\ref{sec:methods}), which could introduce disparities whose resolution is beyond the scope of this work. \citet{Dissauer2018b} performed a statistical analysis on 62 dimming events occurring between 2010 and 2012 and found within $40^\circ$ from the central meridian. The dimmings were associated with flares ranging from B- to X-class, and every event was accompanied by a CME. In Fig.~\ref{fig:cor} grey crosses represent data adapted from the study by \citet{Dissauer2018b}, while blue markers correspond to the May 2024 events analysed in this study. The black (blue) line shows the linear regression fit to the distribution of the \citet[May 2024]{Dissauer2018b} data in log–log space, following $\log{Y}=k\cdot\log{X}+d$, where $k$ and $d$ denote the regression coefficients. We calculate the corresponding correlation coefficients using a bootstrapping method in which N-out-of-N random data pairs are sampled with replacement to recompute $c$ $\approx$10~000 times, and used to extract the mean and standard deviation $\bar{c}\pm\Delta c$. These coefficients ($k$, $d$ and $c=\bar{c}\pm\Delta c$) are indicated in grey (blue) within the panels for the \citet[May 2024]{Dissauer2018b} datasets. We note that we recalculated the regression and correlation coefficients for \citet{Dissauer2018b} data, as in the original study the flare ribbon parameters are considered as functions of the dimming parameters, whereas we compared them the opposite way; that is, the dimming parameter distribution with respect to the flare ribbon parameters.

Figure~\ref{fig:cor}a compares the magnetic dimming area $A_\phi$ against the flare ribbon area $A_{rbn}$. The May 2024 events have a strong correlation of $c=0.65\pm 0.13$, which is in line with the correlation found by \citet{Dissauer2018b} of $c=0.62\pm 0.09$. Overall, the May 2024 events dimming magnetic areas $A_\phi$ are on the upper range of the areas found in the general population studied by \citet{Dissauer2018b}, which is due to the exceptionally large size of AR~13664. Considering the different GOES class of the May 2024 flares (crosses represent X-class flares and dots M-class flares) reveals that stronger flares align better with the prediction from the general population (black line) and show a steeper increase in $A_\phi$ with $A_{rbn}$, whereas weaker flares have a less pronounced slope. 

Figure~\ref{fig:cor}b shows the relationship between the total unsigned magnetic flux $\phi$ of coronal dimmings and flare ribbon reconnection flux $\phi_{rbn}$. We note that we derive the dimming fluxes over the entire dimming area, which encompass both core and secondary dimming regions, following the same approach as \citet{Dissauer2018b}. The methods and resulting quantities are therefore directly comparable. We obtained a correlation of $c=0.60\pm0.18$, which supports the findings from \citet{Dissauer2018b} indicating that the more flux is reconnected in the flare ribbons the more magnetic flux is involved in the dimmings. \citet{Dissauer2018b} reported that strong flares (>M1.0) had a better balance between secondary dimming flux and flare ribbon flux than weaker flares, where the dimming flux is larger than the flare reconnection flux. The solid and dashed black lines indicate this regime where the ratio between fluxes is within 0.5 to 2.0 and thus are considered as roughly balanced. For the May 2024 events we found that X-class flares (blue crosses) contained similar amounts of reconnection flux and dimming flux, while most of the M-class flares (blue dots) had larger dimming flux than flare ribbon flux. Notably, all events after May 9 had reconnection fluxes comparable to the dimming fluxes, while, among the events between May 5 and May 8, the first X-class flare on May 8 was the only flare where the fluxes were comparable. To assess whether this trend could arise by a decrease in the correspondence of the LOS magnetic field to the radial magnetic field in events occurring closer to the limb, we separately calculated the dimming flux using $B_r$ instead of $B_\text{LOS}$. The same trend previously found persisted under this approach.

\subsection{Evolution of coronal dimming location and morphology} \label{sec:location}

\begin{figure*}[h!]
    \centering
    \resizebox{\hsize}{!}{\includegraphics{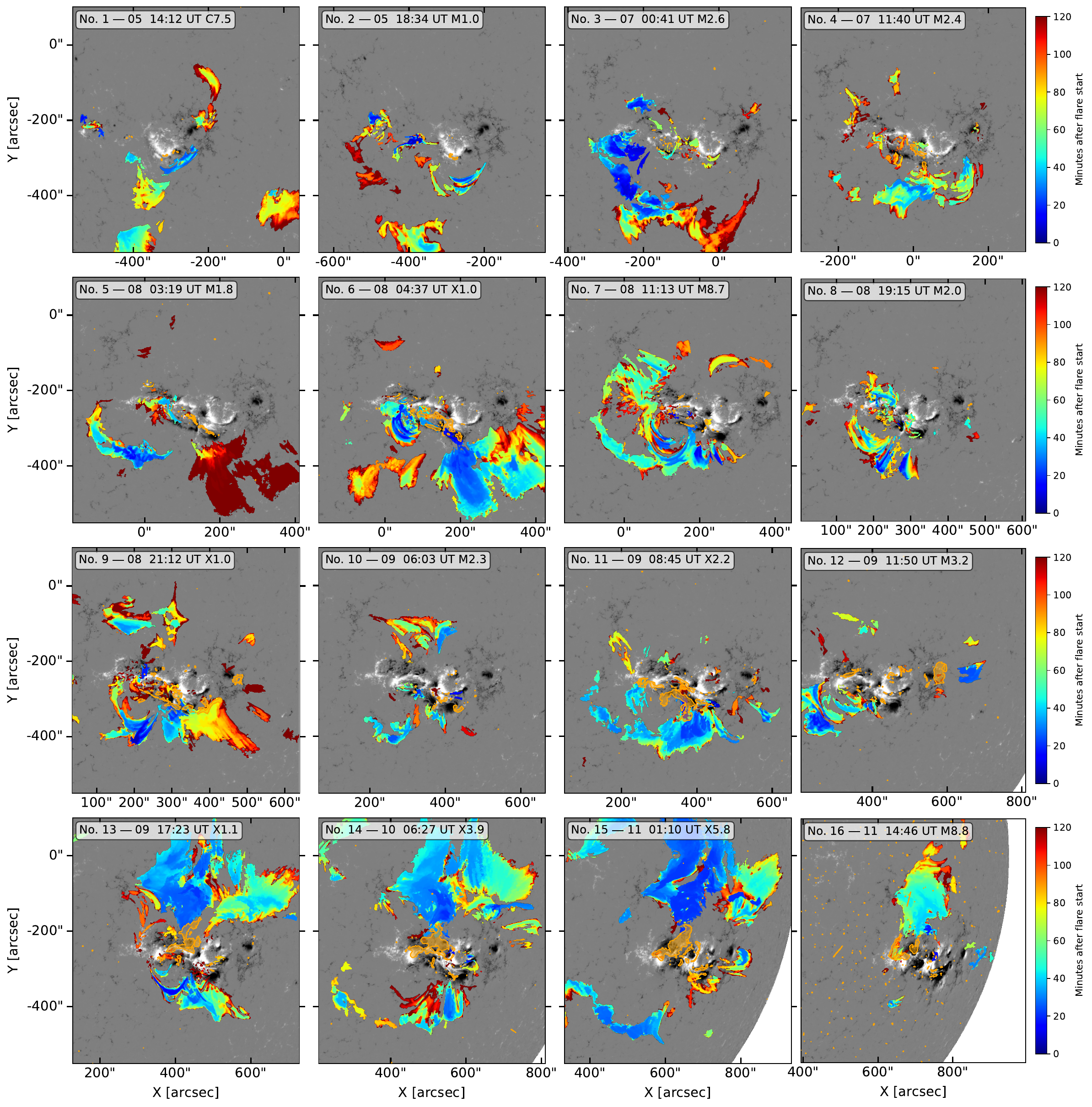}}
    \caption{Timing maps of all coronal dimming events overlaid on HMI LOS magnetograms taken 15~min before the associated flares. The colour of the shading indicates the first detection time of the dimming region in minutes after the flare onset. The flare ribbons detected before the flare end time are shaded in orange. The number indicated in the upper left corner of each subplot is the event number on Table~\ref{table:dimming_properties}.}
    \label{fig:evolution_all}
\end{figure*}

We examined the spatial distribution and morphology of coronal dimmings in AR 13664 over a two-week period. Figure~\ref{fig:evolution_all} presents the timing map of the dimming regions (blue to red) and the flare ribbons (orange) as shaded areas overlaid on the corresponding magnetograms. The colour of the timing map of the dimming regions shows the detection time of the dimming from the flare onset time, as given by the colour bars on the right-hand side. Two distinct morphological groups can be identified: dimmings that expand predominantly southwards and those that expand predominantly northwards. In Tab.~\ref{table:dimming_properties}, the last column indicates the dimming expansion direction for all events under study. When the dimming occurs in both directions, the predominant expansion is listed first. A transition in behaviour occurs after the X1.1 flare on 9 May at 17:23~UT (event no.~13), with all subsequent dimmings expanding primarily northwards, while earlier events expand mainly southwards, except for event no.~10. 

\begin{figure*}[h!]
    \centering
    \resizebox{\hsize}{!}{\includegraphics{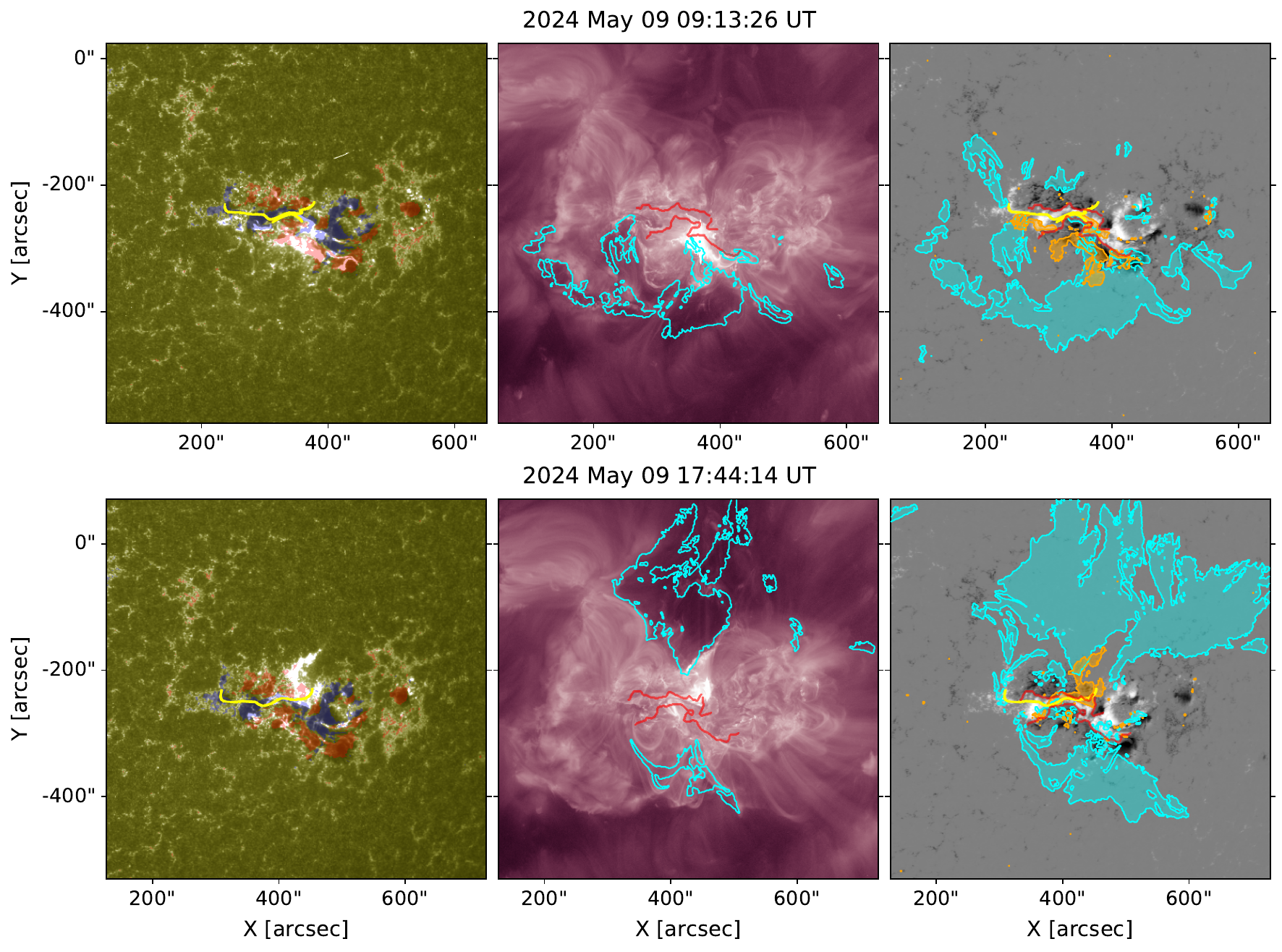}}
        \caption{Overview of the coronal dimming and flare ribbon morphology for event no.~11 (X2.2 flare, May 9 08:45~UT; top row) and event no.~13 (X1.1 flare, 9 May 17:23~UT; bottom row). Left column: AIA 1600~\AA~images at the flare peak. The magnetic field regions $|B|\ge 250$~G are shaded in blue (positive) and red (negative). Middle column: AIA 211~\AA~images at the same time step as in the left column with detected instantaneous dimming contours (cyan). Right column: HMI magnetograms overlaid with the final dimming regions (cyan) and flare ribbons (orange). Red lines show the two main PILs, and the yellow line shows a separatrix layer extracted from squashing factor maps by \citet{Jarolim2024magnetic}. The associated movie is available online.}
        \label{fig:evolution_some}
\end{figure*}

To investigate this morphological shift, we focused on two representative cases: event no.~11 (X2.2 flare, May 9 08:45~UT) and event no.~13 (X1.1 flare, May 9 17:23~UT). The dimming associated with event no.~11 exhibits a pronounced southward expansion with little to no northward component, whereas the dimming in event no.~13 expands primarily northwards, with a dimming region later opening to the south. These cases illustrate the two morphological groups and mark the transition in the predominant direction of dimming expansion between the events. Figure~\ref{fig:evolution_some} presents a comparison of the two events: The first column shows AIA 1600~\AA~images at flare peak with strong magnetic field regions shaded (blue: positive, red: negative), the second column shows AIA 211~\AA~images with the detected dimming contours (cyan) at the same time step, and the third column shows HMI magnetograms with the total dimming regions (cyan) and flare ribbons (orange) shaded. A movie of this figure showing the evolution of the coronal dimming region alongside the flare ribbons is available as an online movie.

At the time of these events, the magnetic configuration of AR~13664 consisted of two main polarity inversion lines (PILs), separated by a narrow, strong positive polarity band extending horizontally through the AR. The two PILs are shown in red in Fig.~\ref{fig:evolution_some}. In event no.~11, the main flare ribbons were located on both sides of the southern PIL, whereas in event no.~13 the ribbons occurred on both sides of the northern PIL, with additional ribbons detected in the southern negative polarity. The flare loops associated with these southern ribbons in event no.~13 are visible in the 211~Å~images (Fig.~\ref{fig:evolution_some}, second column), connecting across the southern PIL. In contrast, no ribbons are observed on the northern negative polarity during event no.~11.

Combining flare ribbon locations with the dimming evolution, we find a consistent relation: flare ribbons located on the northern PIL area associated with dimmings showing a northward expansion, while ribbons on the southern PIL correspond to dimmings expanding southwards. In cases where reconnection involves both PILs, dimmings expand in both directions, with the main ribbons indicating the primary expansion of the dimming region, as exemplified by event no.~13. This relationship is consistently observed in flares between May 8 and May 11 (Fig.~\ref{fig:evolution_all}). Specifically, event no.~5, 6, and 11 display exclusively southward dimming expansion (southern PIL ribbons); event no.~7 and 9 show predominantly southward dimming expansion (southern PIL ribbons with additional ribbons on the northern PIL); events no.~13, 14, and 15 exhibit primarily northward expansion (northern PIL ribbons with smaller ribbons on the southern PIL); and event no.~16 shows exclusively northward expansion (northern PIL ribbons). For events no.~1–4, this description does not apply, as the two major PILs were not yet fully developed. 

These observations indicate that the coronal dimming morphologies may reflect the presence of two distinct magnetic domains, separated by a separatrix layer located approximately above the narrow positive polarity region. This separatrix layer was found by \citet{Jarolim2024magnetic} in NLFF extrapolations of the magnetic field for event no.~9 and 14. We show the separatrix layer taken from squashing factor maps from \citet{Jarolim2024magnetic} for events no.~11 and 13 as a yellow line in Fig.~\ref{fig:evolution_some}. The main flare ribbons occur south and north of the separatrix layer for event no.~11 and no.~13, respectively, as shown in the first and third column in Fig.~\ref{fig:evolution_some}. Thus, reconnection occurs in two different magnetic domains for each flare.

In addition, we find a clear distinction in the tilt of the flare ribbons with respect to the solar equator, with the flare ribbons along the southern PIL tilted southwestwards and the ribbons along the northern PIL tilted northwestwards. The ribbon tilt reflects the orientation of the underlying PIL and thus it may serve as a solar proxy for the flux rope type of the CMEs detected in situ \citep{palmerio2017determining}. \citet{liu2024pileup} associated the geoffectiveness of the May 2024 events CMEs with a southward tilt in the flare ribbons; that is, with events no.~6, 7, 9, and 11. These events also exhibit a predominantly southward dimming expansion.

\subsection{Magnetic structures involved in the eruptions}\label{sec:structures}
\begin{figure*}[t]
    \centering
    \resizebox{\hsize}{!}{\includegraphics{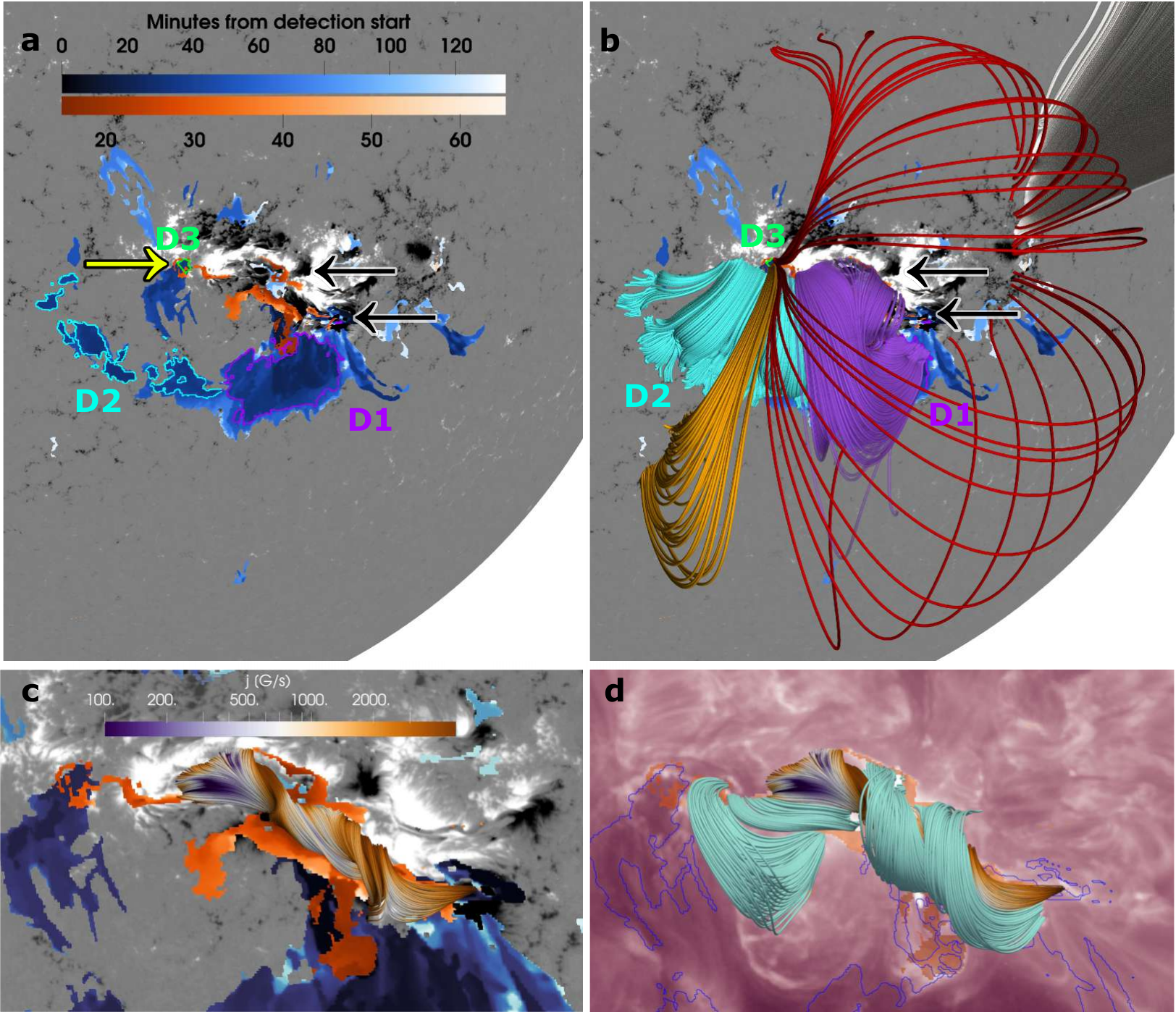}}
        \caption{Magnetic field structures rooted in the coronal dimmings of event no.~11 overlaid on an HMI radial magnetogram taken on May 9 at 08:36~UT.  Coronal dimming regions are shaded in blue and flare ribbons in orange. Black and yellow arrows point to the flare ribbons. (a) Subregions of dimmings contoured in purple, cyan, and green and denoted as D1-D3. (b) All magnetic field structures from the PFSS extrapolation from D1 in purple, D2 in cyan, and D3 in orange and red. Grey field lines show open flux near the AR. (c) and (d) Field lines sampled from NLFF extrapolations showing the flux rope (panel a, colour refers to local current density) and the field lines connected to the flare ribbons (panel d, cyan).}
        \label{fig:app:pfss_example}
\end{figure*}

\begin{figure*}[t]
    \centering
    \resizebox{\hsize}{!}{\includegraphics{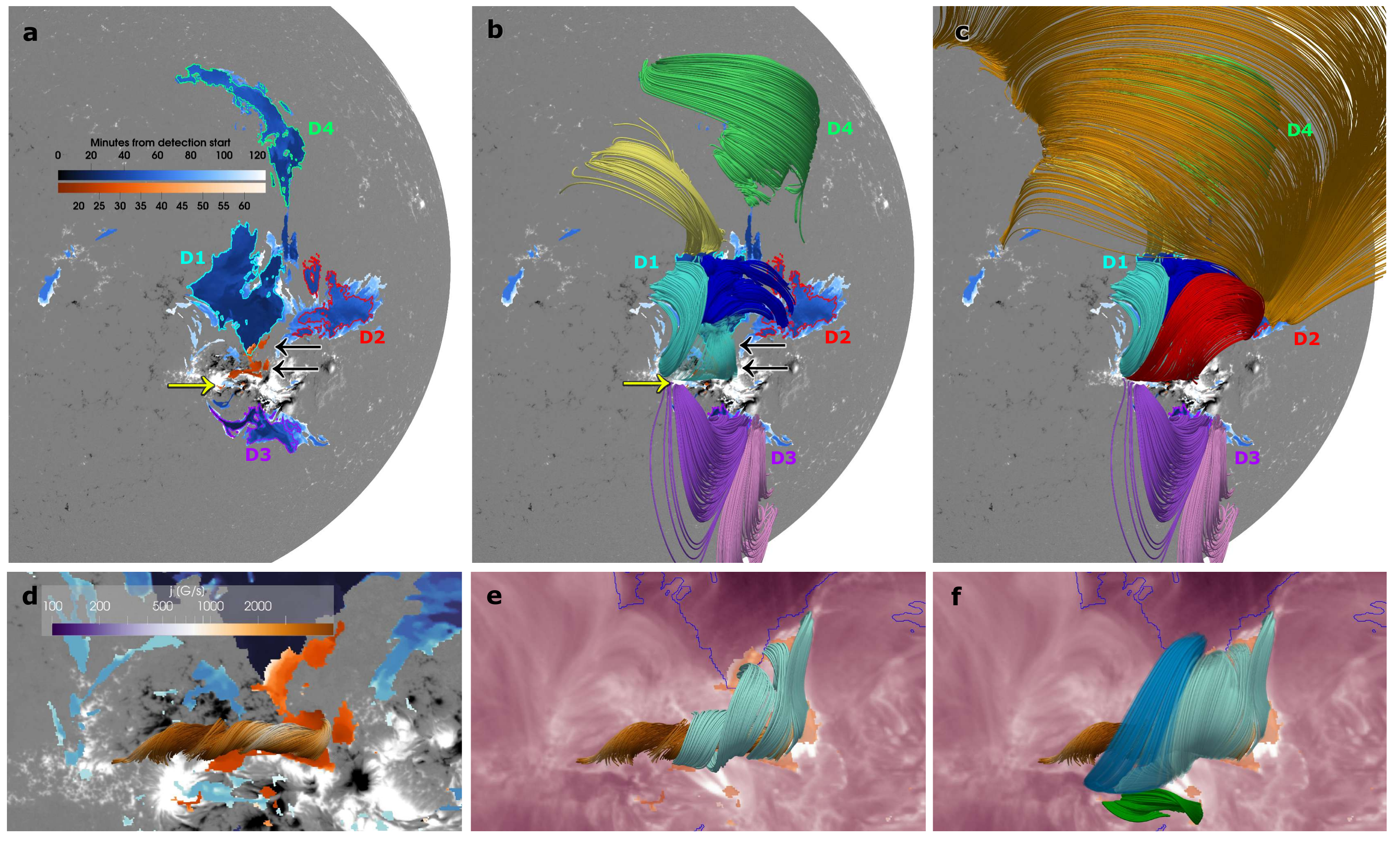}}
        \caption{Same as Fig.~\ref{fig:app:pfss_example} but for event no.~13. (a) Subregions of dimmings contoured in purple, red, cyan, and green and denoted as D1-D4. (b) and (c) Magnetic flux systems from the PFSS extrapolation from D1, D3, and D4 (panel b) and from all dimming regions (panel c). (d) and (f) Field lines sampled from NLFF extrapolations showing the flux rope (panel~d, the colour indicates the local current density) and the field lines connected to the flare ribbons (panels e and f in cyan, blue and green).}
        \label{fig:app:pfss_example_2}
\end{figure*}

Figures~\ref{fig:app:pfss_example} and \ref{fig:app:pfss_example_2} show the PFSS and NLFF extrapolations for event no.~11 (X2.2 class flare on May 9 at 08:45~UT), and no.~13 (X1.1 flare on May 9 at 17:23~UT), respectively. We employed the NLFF extrapolations by \citet{Jarolim2024magnetic} at 08:36~UT and 17:12~UT on May 9. These two events exhibit representative morphological features of the southwards expanding dimmings (event no.~11) and northwards expanding dimmings (event no.~13). Our objective is to examine how the magnetic flux systems involved in these eruptions differ between the two dimming types, and to characterise them following the classification framework proposed by \citet{Veronig2025}. Figures~\ref{fig:app:first_flare} and \ref{fig:app:second_flare} and their accompanying movies give an overview of event no.~11 and 13, respectively, showing large field of view AIA~211~\AA~images, the base-ratio equivalent, and AIA~131, 171, 304~\AA~images of the AR during the events. Notably, at the time of these events, a large coronal hole (CH) was located in the northern hemisphere and a smaller CH to the south of the AR, both positioned eastward of the AR. 

\begin{table*}
    \caption{\label{table:dimming_subregions}Properties of the coronal dimming subregions in events no.~11 and 13.}
    \centering
    \begin{tabular}{cccccccc}         
    \hline\hline          
    N & Start & Class & Dimming & \CellWithForceBreak{$A_\phi$ \\ ($10^{9} \text{km}^2$)} &  \CellWithForceBreak{$\phi$ \\ ($10^{20} \text{Mx}$)} & \CellWithForceBreak{$B$ \\ (G)} & Dimming classification \\ 
    \hline
    \multirow{3}{*}{11} & \multirow{3}{*}{08:45} & \multirow{3}{*}{X2.2} & D1 & $ 3.3 $ & $ 16.5 $ & 102 & Strapping flux dimming \\
    &  &  & D2 & $ 0.9 $ & $ 1.9 $ & 41 & Exterior dimming \\
    &  &  & D3 & $ 0.2 $ & $ 2.5 $ & 234 & Exterior and open-flux dimming \\
    \hline
    \multirow{4}{*}{13} & \multirow{4}{*}{17:23} & \multirow{4}{*}{X1.1} & D1 & $ 2.9 $ & $ 5.9 $ & 40 & Strapping flux and exterior dimming \\
    &  &  & D2 & $ 2.8 $ & $ 6.3 $ & 45 & Strapping flux, exterior, and potentially open-flux  dimming \\ 
    &  &  & D3 & $ 2.2 $ & $ 8.7 $ & 80 & Exterior dimming \\
    &  &  & D4 & $ 2.4 $ & $ 3.1 $ & 26 & EUV wave triggered dimming \\
    \hline 
    \end{tabular}
    \tablefoot{We list the start time on May 9 and the class of the associated flare derived from the GOES flare catalogue. We list the dimming name, magnetic dimming area ($A_\phi$), the total unsigned magnetic flux of the dimming ($\phi$), and the mean magnetic flux density ($B$). In the `Dimming classification' column we list the dimming type based on the classification framework by \citet{Veronig2025}, as described in Sect.~\ref{sec:structures}.}
\end{table*}

Figures~\ref{fig:app:pfss_example}a and \ref{fig:app:pfss_example}b illustrate the PFSS extrapolation for event no.~11 overlaid on the HMI radial magnetogram taken on May 9 at 08:36~UT, 9~minutes before the flare start time. In Fig.~\ref{fig:app:pfss_example}a we show the coronal dimming in blue shading and the flare ribbon in orange shading, where darker regions correspond to earlier detection times. The flare ribbons are located on both sides of the southern PIL. Black arrows indicate the main flare ribbons, and a yellow arrow marks a circular flare ribbon region located east of the PIL. We consider dimming regions detected within 30~min from the flare onset and manually identify three distinct subregions, labelled D1-D3, outlined in purple, cyan, and green contours in Fig.~\ref{fig:app:pfss_example}a, respectively. D1 represents the main dimming region, occurring earliest and exhibiting the strongest intensity decrease (see Fig.~\ref{fig:methods:dimming}c–d). D2 appears as an elongated dimming region that expands from D1 eastwards and shows a shallower intensity decrease. D3 is an isolated dimming region mainly selected by its location within the circular flare ribbon region indicated by the yellow arrow.  

We computed the unsigned magnetic flux ($\phi$), mean magnetic field density ($B$), magnetic area ($A_\phi$), and polarity for each subregion, which are summarised in Table~\ref{table:dimming_subregions} alongside their respective dimming classification type. Subregions D1 and D2 exhibit negative polarity, consistent with their connection to the central positive‐polarity band of the AR, whereas D3 displays positive polarity. Among the three, D1 is the largest in both area and magnetic flux, with $A_\phi$ of $3.3~\times10^{9}$~km$^{2}$ and unsigned magnetic flux $\phi$ of $16.5\times10^{20}$~Mx. Subregions D2 and D3 are considerably smaller: D2 has $A_\phi=0.9~\times10^{9}$~km$^{2}$ and $\phi=1.9\times10^{20}$~Mx, while D3 has $A_\phi=0.2~\times10^{9}$~km$^{2}$ and $\phi=2.5\times10^{20}$~Mx. These values further evidence that D1 is the dominant contributor to the overall dimming. Regarding the mean magnetic field density, D3 exhibits the strongest field ($B = 234$~G), reflecting its location in the AR core, followed by D1 ($B=102$~G) and D2 ($B=41$~G). These three regions are used as source regions for the field lines shown in panel b of Fig.~\ref{fig:app:pfss_example}.

The extrapolated PFSS magnetic flux systems corresponding to each dimming subregion are shown in Fig.~\ref{fig:app:pfss_example}b: purple field lines are rooted in D1, cyan field lines in D2, and red and orange field lines in D3. Field lines from D1 primarily connect the dimming region with the positive polarity band, arching above the PIL, as indicated by the black arrows marking the flare ribbons located beneath these field lines. Moreover, D1 occurs almost co-temporally to the flare ribbons, as it can be observed in the movie accompanying Fig.~\ref{fig:evolution_some}. Hence, this magnetic flux system is classified as strapping flux. Strapping flux dimmings occur when the erupting flux rope is expanding and the strapping flux is lifted, which is observed as an outwards expanding dimming \citep{Veronig2025}. 

Field lines rooted in D2 (cyan) connect the curved dimming region to a weaker positive-polarity patch located beneath the circular flare ribbon indicated by the yellow arrow in Fig.~\ref{fig:app:pfss_example}a. Other D2 field lines are low-lying and connect locally south of the dimming region. This configuration suggests that D2 is an exterior dimming \citep{Veronig2025}. Exterior dimmings occur when the erupting structure interacts with adjacent closed magnetic flux systems, leading to plasma evacuation through either reconnection or the upward expansion of the neighbouring flux. In event no.~11, reconnection around D3 induces partial reconnection and expansion of the field lines originating in D2, which in turn triggers localised reconnection at D2 itself. The resulting brightenings around D2 are visible in other AIA channels, such as 131 and 304~\AA~(see Fig.~\ref{fig:app:first_flare}). In addition, this dimming occurs after the primary dimming at D1, as shown in the timing maps (Fig.~\ref{fig:methods:dimming}c), further supporting its interpretation as an exterior dimming. Notably, this curved dimming feature is found in other four events (no.~5, 7, 10, and 11), as observed in Fig.~\ref{fig:evolution_all}.

Lastly, the field lines rooted in D3 can be divided into two flux systems shown in red and orange in Fig.~\ref{fig:app:pfss_example}b: orange field lines connect to the boundary of the CH south-east of the AR, and red field lines vault over the AR and connect to the strong negative-polarity sunspot west of the AR. The later field lines (red) closely align with an open field region shown in grey in Fig.~\ref{fig:app:pfss_example}b. The closeness of the flux systems to open field lines suggest that D3 could represent a combination of an exterior dimming with an open-flux dimming, which occur when erupting flux reconnects with open flux \citep{Veronig2025}. The interpretation as an open-flux dimming is supported by the brightenings observed near the negative polarity sunspot west of the AR, as shown in orange in the top right panel of Fig.~\ref{fig:evolution_some} and in Fig.~\ref{fig:app:pfss_example}a, and in the movie from Fig.~\ref{fig:app:first_flare}.

Figure~\ref{fig:app:pfss_example}c shows the flux rope derived from NLFF extrapolations extending northeast to southwest above the southern PIL. The western footpoint lies close to the D1 dimming, whereas the eastern footpoint is located west of D3 and has no neighbouring dimming regions. Figure~\ref{fig:app:pfss_example}d shows field lines traced from selected flare ribbon regions; that is, field lines that are expected to eventually reconnect as the flux rope expands. We note that some of these field lines link the main flare ribbons to the elongated ribbon between the flux rope and the circular ribbon around D3 (yellow arrow in Fig.~\ref{fig:app:pfss_example}a). These lines may have undergone reconnection, extending the eruption towards D3 and producing the exterior and open-flux dimming previously described at D2. 

Figures~\ref{fig:app:pfss_example_2}a-c present the PFSS extrapolation for event no.~13 overlaid on top of the HMI radial magnetogram taken on May 9 at 17:12~UT, 11~minutes before the flare onset. In Fig.~\ref{fig:app:pfss_example_2}a we show the coronal dimming expanding to the northern hemisphere, and the flare ribbons located along the northern PIL. Black arrows point to the main flare ribbons, while a yellow arrow marks a small flare ribbon region below the positive polarity band. We analysed the dimming regions detected within the first 45~min after flare onset and divide the coronal dimming into four distinct subregions, labelled D1-D4, outlined in cyan, red, purple, and green contours, respectively. Figure~\ref{fig:app:pfss_example_2}c shows the magnetic flux systems connected to these coronal dimming subregions: cyan, blue and yellow field lines are rooted in D1, red and orange field lines in D2, purple and pink ones in D3, and green field lines in D4. 

The dimming regions D1-D3 exhibit negative polarity, while D4 has positive polarity. The largest regions are D1 and D2, with a magnetic area $A_\phi$ of $2.9\times10^9$~km$^{2}$ and $2.8\times10^9$~km$^{2}$, respectively (see Table~\ref{table:dimming_subregions}). In comparison, D3 and D4 have smaller areas of $A_\phi=2.2\times10^9$~km$^{2}$ and $A_\phi=2.4\times10^9$~km$^{2}$. Dimming subregion D3 carries the largest magnetic flux at $8.7\times10^{20}$~Mx. However, the combined unsigned magnetic flux of D1 and D2 reaches $12.2\times10^{20}$~Mx, making these two regions main contributors to the overall dimming; D4, in contrast, contains only $\phi=3.1\times10^{20}$~Mx. With respect to the mean magnetic field density, D3 lies closer to the AR core and therefore contains stronger fields, with $B=80$~G. Subregions D1 and D2 exhibit mean field strengths of $B=40$~G and $B=45$~G, respectively. Region D4, on the other hand, contains a mean magnetic field density of 26~G, only slightly above the noise level of HMI.

To further characterise the magnetic flux systems, Fig.~\ref{fig:app:pfss_example_2}b provides a view of the AR showing only the field lines associated with the dimmings D1, D3 and D4. The cyan field lines rooted in D1 are strapping field lines, arching above the northern PIL and connecting to the southern flare ribbons marked by the black arrows. 
The dark blue field lines are not directly connected to the AR, though some connect to D2, suggesting that this flux system may participate in the eruption through reconnection. Accordingly, this represents an exterior dimming. The yellow field lines rooted at the northern tip of D1 connect with the boundary of the northern CH, as shown in Fig.~\ref{fig:app:pfss_example_2}b. This flux system likely remains uninvolved in the eruption, and the observed extent of the dimming region is most likely determined by the stretching of the overlying fields and bending of the yellow field lines (see movie from Fig.~\ref{fig:app:second_flare}). Thus, D1 is a combination of both strapping-flux and exterior dimming. Moreover, in Fig.~\ref{fig:app:pfss_example_2}a we can observe the northern flare ribbon sweeping across D1 (light orange colour over the D1 cyan boundary), which indicates that part of the strapping flux was added to the flux rope via strapping–strapping reconnection. In this scenario, as the flare reconnection takes field lines further from the PIL, the strapping field is also swept by the flare ribbons.

Field lines rooted in D2 are shown in Fig.~\ref{fig:app:pfss_example_2}c in red and orange colours. We identify two distinct flux systems: the red flux system connects to the central positive polarity band below the southern ribbon, while the orange flux system links D2 to the boundary of the northern CH (see Fig.~\ref{fig:app:second_flare}). Therefore, D2 is a combination of strapping-flux dimming and exterior dimming. The dark blue flux system rooted in D1 may also contribute to the expansion of D2. Reconnection and stretching of the strapping (red) and exterior flux (dark blue) may lead to the stretching and potential reconnection of the exterior field (orange), accounting for the final extent of the D2 dimming. Similar to event no.~11, D2 might also include an open-flux dimming since some of the orange field lines are open (extend to the source surface height). Although we do not observe signs of reconnection, this open flux might have nevertheless participated in the eruption.

The dimming morphologies of events no.~13–15 exhibit strong similarities, as shown in Fig.~\ref{fig:evolution_all}. In all three cases, the northern dimming regions (corresponding to D1 and D2 in Fig.~\ref{fig:app:pfss_example_2}) evolve comparably: D1 forms first, followed shortly by the development of D2, after which the flare ribbons sweep across D1. This consistent sequence indicates that the same underlying magnetic flux systems are likely involved in all three eruptions.

Although the primary flare ribbons are located along the northern PIL, some reconnection also occurs along the southern PIL, as indicated by the yellow arrow in Fig.~\ref{fig:app:pfss_example_2}a and discussed in Sect.~\ref{sec:location}. These ribbons are associated with flux systems rooted in D3, which are shown in purple and pink in Fig.~\ref{fig:app:pfss_example_2}b-c. The purple field lines connect D3 with the southern flare ribbons, indicating that D3 is also an exterior dimming, whereas the pink field lines appear to be unrelated to the eruption. 

Figure~\ref{fig:app:pfss_example_2}d presents the flux rope of event no.~13 from NLFF extrapolations. The structure is long, highly twisted, and positioned closer to the southern ribbon, resulting in an asymmetric flux rope and ribbon configuration. In Fig.~\ref{fig:app:pfss_example_2}e we show representative field lines connecting the two principal flare ribbons. The western footpoint of the flux rope is embedded in field lines linking to the northernmost ribbon segment, where reconnection starts. The blue field lines in Fig.~\ref{fig:app:pfss_example_2}f represent those that reconnect as the flare ribbons sweep across the northern dimming region. The two small flare ribbons south of the main reconnection site are connected by the green field lines in Fig.~\ref{fig:app:pfss_example_2}f. In Figs.~\ref{fig:app:pfss_example_2}e-f we can observe the post flare loops bridging the central positive polarity band and the small southwestern ribbon. This reconnection might have facilitated the reconfiguration of the green field lines linked to the southeastern ribbon. The latter ribbon region, indicated by the yellow arrow in Figs.~\ref{fig:app:pfss_example_2}a-b, is magnetically connected to the aforementioned D3 dimming region.

At the full-disc scale, an elongated dimming region (D4) extending northwards from the AR is observed (green contours in Fig.~\ref{fig:app:pfss_example_2}a). We consider three possible explanations for its formation: 1) projection effects making a pre-existing low-density region visible after the expansion of the overlying field lines (indicated in orange), 2) large-scale magnetic connectivity enabling reconnection to occur in the northern hemisphere, or 3) reconnection triggered by the coronal EUV wave associated with the CME. The PFSS extrapolation reveals no direct magnetic connection between D4 and the AR (green field lines in Fig.~\ref{fig:app:pfss_example_2}b), which makes the large-scale connectivity explanation unlikely. We note that overlying field lines above D4 also showed no connectivity with the AR nor the other dimming regions. However, AIA~211~\AA~observations show local brightenings around D4 (see movie from Fig.~\ref{fig:app:second_flare}), consistent with reconnection signatures, suggesting that projection effects alone might not cause the dimming. During this event, an EUV wave can be observed, as shown in Fig.~\ref{fig:app:second_flare} and the accompanying movie. The temporal and spatial evolution of D4 closely tracks the propagation of the wave front, supporting the interpretation that the wave perturbs the local magnetic field, triggering reconnection and producing a density depletion, similar to the case discussed in \citet{zhou2020magnetic}. Moreover, the proximity of D4 to the small CH north from the AR may facilitate flux opening and contribute to the observed dimming evolution.

\section{Summary and discussion} \label{sec:discussion}
We have presented an analysis on the coronal dimmings associated with the May 2024 solar energetic events from AR~13664 using SDO/AIA and HMI data. We investigated the spatial extent and morphology of 16 coronal dimmings alongside their accompanying flare ribbons and how they evolved in time. For two representative dimming events, we used PFSS and NLFF extrapolations of the magnetic field to analyse which magnetic flux systems are involved in the eruptions. In the following, we summarise the most important findings:
\begin{enumerate}
    \item The coronal dimmings from AR~13664 exhibited a predominant southward expansion before May 9 and a predominantly northward expansion thereafter. This transition is consistent with a change in the flare ribbon locations with respect to the two major PILs of AR~13664. Events with ribbons rooted close to the southern PIL were accompanied by southwards directed dimmings, whereas events with ribbons rooted close to the northern PIL were associated with northward expanding dimmings.
    \item The PFSS extrapolations of events no.~11 and 13 reveal that southwards and northwards expanding dimmings originate from a distinct magnetic flux system. Both dimmings are mainly formed as strapping-flux dimmings, with the strapping fluxes arching over the two distinct PILs. Exterior dimmings are also formed in the predominant dimming expansion direction, as flux external to the AR takes part in the eruption.
    \item The flare ribbon area $A_{rbn}$ for the May 2024 events has a strong correlation with the magnetic area of the coronal dimmings $A_\phi$ ($c=0.65\pm0.13$). Similarly, the flare reconnection flux $\phi_{rbn}$ has a strong correlation with the unsigned magnetic dimming flux $\phi$ ($c=0.60\pm0.18$). Notably, the two flux types are more similar for flares after May 8.
\end{enumerate}

\subsection{Magnetic domains of AR~13664}
The sequence of events from AR~13664 reveals that the combined information of the location of flare ribbons and coronal dimmings provides insight into the existence of different magnetic domains in the AR. Throughout the first passage of the AR over the solar disc, it exhibited two strong primary PILs, which were roughly horizontal and separated by a positive polarity band. 
The dimming expansion direction evidenced the existence of two magnetic domains, each containing one of the primary PILs. We showed that reconnection occurring above the northern PIL produced dimmings expanding northwards, while reconnection above the southern PIL generated dimmings expanding southwards. Thus, the flux systems that eventually took part in the eruptions were confined to the magnetic domain where the initial reconnection took place in (associated with either the northern or southern PIL). The existence of these two domains predicted by the dimming expansion is in agreement with the findings from NLFF modelling by \citet{Jarolim2024magnetic}, who showed a major separatrix layer, i.e. a topological boundary dividing magnetic domains, running along the central positive polarity region from May 8. 

\subsection{Magnetic flux systems involved in the eruptions}
Coronal dimmings are the most prominent low-corona signatures of CMEs and therefore provide a key diagnostic for identifying the magnetic flux systems involved in the eruptions and the physical processes that govern their evolution (see the categorisation and examples in \citealt{Veronig2025}). To exploit this diagnostic potential, we combined PFSS and NLFF extrapolations of the coronal magnetic field with flare ribbon and coronal dimming detections. We performed magnetic field extrapolations for two events (no.~11 and 13) with different representative features of the coronal dimmings in the May 2024 events.  We identified multiple dimming subregions in both events, and we classified them following the dimming characterisation framework by \citet{Veronig2025}.

For events no.~11 and 13, the main dimming regions (D1) correspond to a strapping-flux dimming overlying the respective PIL next to where the flare ribbons occur (see Figs.~\ref{fig:app:pfss_example} and \ref{fig:app:pfss_example_2}). The strapping flux is lifted as the flux rope expands, producing the observed density depletion in the form of the dimming signature. In event no.~13, we additionally observed strapping–strapping reconnection as flare ribbons sweep across the coronal dimming region, indicating that part of the strapping flux within D1 is added to the erupting flux rope.  Both events also exhibit exterior dimmings. In event no.~11, secondary reconnection was observed indirectly as small flare ribbons east of the flux rope location, which produced exterior dimmings at D2 and D3 (Fig.~\ref{fig:app:pfss_example}). Notably, D3 also contains an open-flux dimming component, as field lines rooted in D3 reconnect with open field lines to the northwest of the AR.

Exterior dimmings are likewise present in event no.~13 (Fig.~\ref{fig:app:pfss_example_2}). The dimming subregion D3 forms as reconnection occurs south of the main southern flare ribbon, leading to the expansion of magnetic field lines linking the central positive polarity band of the AR with D3. The dimming region D2 represents a combination of strapping-flux and exterior dimming. As the strapping flux is lifted, overlying field lines are stretched and undergo partial reconnection, leading to the final extent of D2. 

The diversity of dimming categories identified in the May 2024 events -- strapping-flux, exterior and open-flux dimmings -- reflects the intrinsic complexity of AR~13664. In addition, individual dimming regions contained several types of dimmings. As eruptions evolve, multiple flux system become involved, making it likely that isolated dimming regions include different flux types.
Moreover, this analysis underscores the utility of the classification framework by \citet{Veronig2025} and shows that through the combination of coronal dimming analysis and magnetic field extrapolations one cannot only identify the magnetic flux systems participating in eruptions, but also the processes by which they become involved in them.

\subsection{Remote opening of flux}
In event no.~13, we detected a large, elongated dimming (D4) extending northwards from the AR  that appears disconnected to the eruption site (Fig.~\ref{fig:app:pfss_example_2}). Its evolution follows the passage of a coronal EUV wave, suggesting that the wave triggered local reconnection and flux opening. Several studies have shown the potential of EUV waves to destabilise remote structures and produce sympathetic flare or eruptions \citep{liu2014advances, zhou2020magnetic}. However, due to the different mechanisms that could couple an EUV wave with the subsequent flare and/or eruption \citep{schrijver2013pathways}, only a few events where this occurs have been found. For event no.~13, the PFSS extrapolations did not show any connection to the flaring AR, and its localised and sharp nature in AIA~211~\AA~observations suggest that projection effects alone cannot produce the observed D4 dimming. Thus, we favour the interpretation that the EUV wave triggered magnetic reconnection that led to the flux being opened, a process likely facilitated by the presence of the adjacent CH.

\subsection{Morphology and evolution of dimmings}
From the analysis of coronal dimming morphology and flare ribbon evolution over a two-week period, we identified similarities in the dimming regions across different events. These patterns suggest that the magnetic flux systems found in the PFSS extrapolations for events no.~11 and 13 are likely involved in other events as well. In particular, the exterior dimming D2 found in event no.~11 is also present in events no.~5, 7, 10, and 11 (see Fig.~\ref{fig:evolution_all}). Likewise, the northern D1 and D2 dimmings (composed of strapping and exterior dimmings) in event no.~13 reappear in events no.~14 and 15. The D4 dimming, generated by the passage of an EUV wave, is similarly detected in these two events, each of which is also accompanied by an EUV wave. Together, these three events form a set of homologous eruptions characterised by flare ribbons laying on both sides of the northern PIL, signatures of strapping-strapping reconnection, and comparable temporal dimming evolution. Events no.~13-15 occur on three separate days (May 9 to May 11), and each is accompanied by a CME. Although the investigation of the mechanisms for the occurrence of homologous flares and CMEs (e.g. \citealt{chertok2004homologous, liu2017causes, suraj2022homologous}) is out of the scope of this study, such processes must have reconfigured the large-scale magnetic field into the configuration observed for event no.~13. 

Furthermore, we found that the flare ribbon location and consequently also the predominant dimming expansion direction change over time. From May 7 to May 9 the flare ribbons are located mostly around the southern PIL, shifting to the northern PIL thereafter. This transition reflects the gradual buildup of energy north of the positive polarity band, which grows as reconnection occurs in the southern PIL, ultimately leading to the destabilisation of the northern magnetic domain. 

\subsection{CME geoeffectiveness}
The different dimming morphologies can also be related to the geoeffectiveness of the associated CMEs. The in situ signatures of the May 2024 geomagnetic storms were studied by \citet{liu2024pileup}. They reported two complex ejecta, i.e. disordered magnetic fields with no flux rope structure \citep{burlaga2001fast}, with differing geoeffectiveness. The first complex ejecta, observed from May 10, contained a strong southward magnetic field component that drove a rapid decrease in the Dst index, reaching values as extreme as $-412$~nT. In contrast, the second ejecta, observed after May 13, was characterised by a comparatively weak southward magnetic component and was associated with slight dips in the recovery of the Dst index. \citet{liu2024pileup} suggested that the first complex ejecta is connected to the CMEs associated with events no.~6, 7, 9, and 11, and the second ejecta is associated with events no.~13-15. A later study by \citet{Weiler2024first} identified the individual ejecta of the CMEs associated with the events no.~6, 7, 9, and 11 within the first complex ejecta. This separation in geoeffectiveness coincides with the predominant coronal dimming expansion of the associated events, with the first complex ejecta associated with dimmings expanding southwards and the second complex ejecta with dimmings expanding northwards.

\citet{wang2024unveiling} examined the flares associated with these complex ejecta using HMI and AIA 131~\AA~observations and suggested that the events no.~6, 7, 9, and 11 had magnetic flux ropes or sheared arcades with southward axial and poloidal fields, while events no.~13-15 had mostly northwards fields. They attributed the differing geoeffectiveness of the resulting ejecta to variations in the orientation of the axial field, consistent with the interpretation by \citet{Weiler2024first}. In our study, we found the same northwards vs southwards tilted division in the flare ribbon detections, and the NLFF extrapolations highlight the existence of these two types of flux ropes. Consequently, the events with the largest dimming regions (events no.~13-15), which are associated with the northwards tilted PILs, did not cause significant geomagnetic disturbances.

\subsection{Flare reconnection fluxes and dimming fluxes}
We found a strong correlation between magnetic dimming area and flare ribbon area in the May 2024 events dimmings ($r\approx0.65$). In addition, the magnetic flux associated with the dimming regions showed a strong correlation with the reconnection flux extracted from the flare ribbons ($r\approx0.60$). This aligns with the findings presented by \citet{Dissauer2018b} and \citet{qiu2017gradual} and reinforces the picture that a substantial amount of poloidal flux is added to the erupting flux rope through reconnection during the eruptions \citep{lin2004role}, and that the overlying expanding and partially reconnecting fields are traced by the secondary dimmings.

\citet{Dissauer2018b} reported a better balance between flare ribbon and coronal dimming magnetic fluxes for $>$~M1.0 flares as compared to weaker flares. In the context of the May 2024 events, we find that X-class flares contained comparable dimming and reconnection fluxes, while most of the M-class flares -- specially before May 8 -- had larger dimming fluxes than reconnection fluxes. On the one hand, this could be due to uncertainties in the detection of flare ribbons, which are larger for weaker flares \citep{Kazachenko2017database}. On the other hand, it could be due to exterior dimmings. Because of the large size and complexity of AR~13664, a significant amount of exterior flux could be lifted, contributing to the dimming flux without directly corresponding to the flare ribbon reconnection flux.

\section{Conclusion} \label{sec:conclusions}
This study provides the first systematic analysis of how the morphology, expansion, and magnetic connectivity of coronal dimmings from AR~13664 relate to flare ribbons, magnetic flux systems involved in the eruptions, and the domains they correspond to. 
Building on the characteristic parameter analysis by \citet{arazquin2025coronal}, we showed the strong correlation of magnetic fluxes and areas between flare ribbons and coronal dimmings, especially in the strongest flares (X-class). 
By tracing the dimming morphology and expansion across successive events, we identified two distinct magnetic domains within the AR' determined the specific flux systems participating in the eruptions, which would otherwise be untraceable' and showed that these systems participate in homologous events. We further showed that the dimming expansion, alongside the orientation of the flare ribbons along the PIL, can explain the different geoeffectiveness of the associated CMEs. In addition, the presence of dimmings far from the AR provide indications of the flux opening produced by large-scale EUV waves initiated by the eruptions. Together, these results underscore the rich potential of coronal dimmings to detect and illuminate the physics behind solar eruptions.

\begin{acknowledgements}
This project has received funding from the European Union's Horizon Europe research and innovation programme under grant agreement No 101134999 (SOLER). The research was sponsored by the DynaSun project and has thus received funding under the Horizon Europe programme of the European Union under grant agreement (no. 101131534). Views and opinions expressed are however those of the author(s) only and do not necessarily reflect those of the European Union and therefore the European Union cannot be held responsible for them. This research was supported by the International Space Science Institute (ISSI) in Bern, through ISSI International Team project no. 516. SDO data are courtesy of NASA/SDO and the AIA, and HMI science teams. GOES is a joint effort of NASA and the National Oceanic and Atmospheric Administration (NOAA). 
\end{acknowledgements}

\bibliographystyle{aa}
\bibliography{My_References}

\begin{appendix}
\onecolumn
\section{Observational overview of events no.~11 and 13} \label{appendix}
\begin{figure*}[h]
    \centering
    \resizebox{\hsize}{!}{\includegraphics{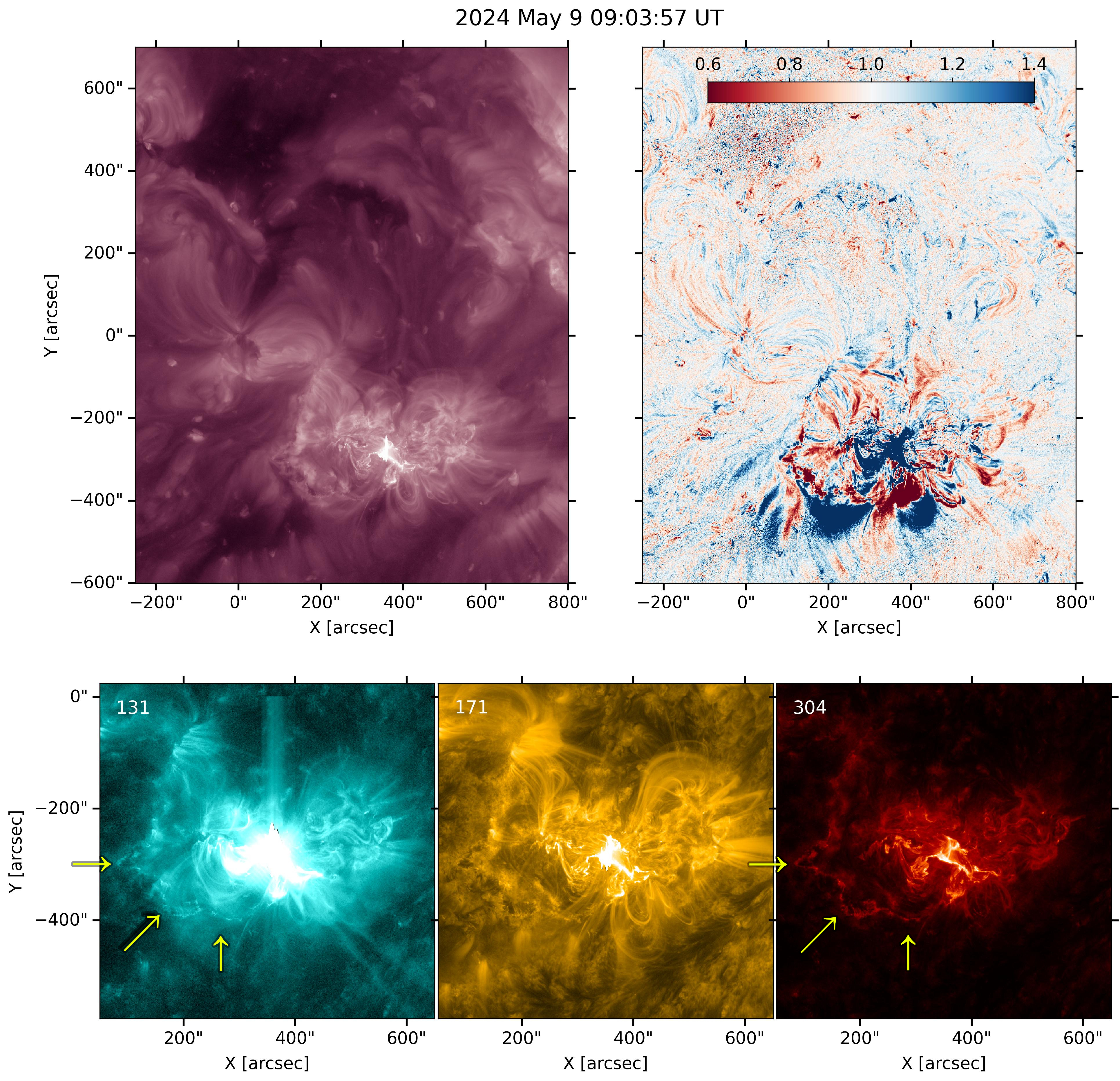}}
        \caption{Overview of the X2.2 flare on 2024 May 9 at 09:28~UT (event no.~11). First row: Direct (left) SDO/AIA 211~\AA~ image and base ratio (right) equivalent. Second row: AIA 131, 171, and 304~\AA~images of AR~13664. The associated movie is available online.}
        \label{fig:app:first_flare}
\end{figure*}

\begin{figure*}[t]
    \centering
    \resizebox{\hsize}{!}{\includegraphics{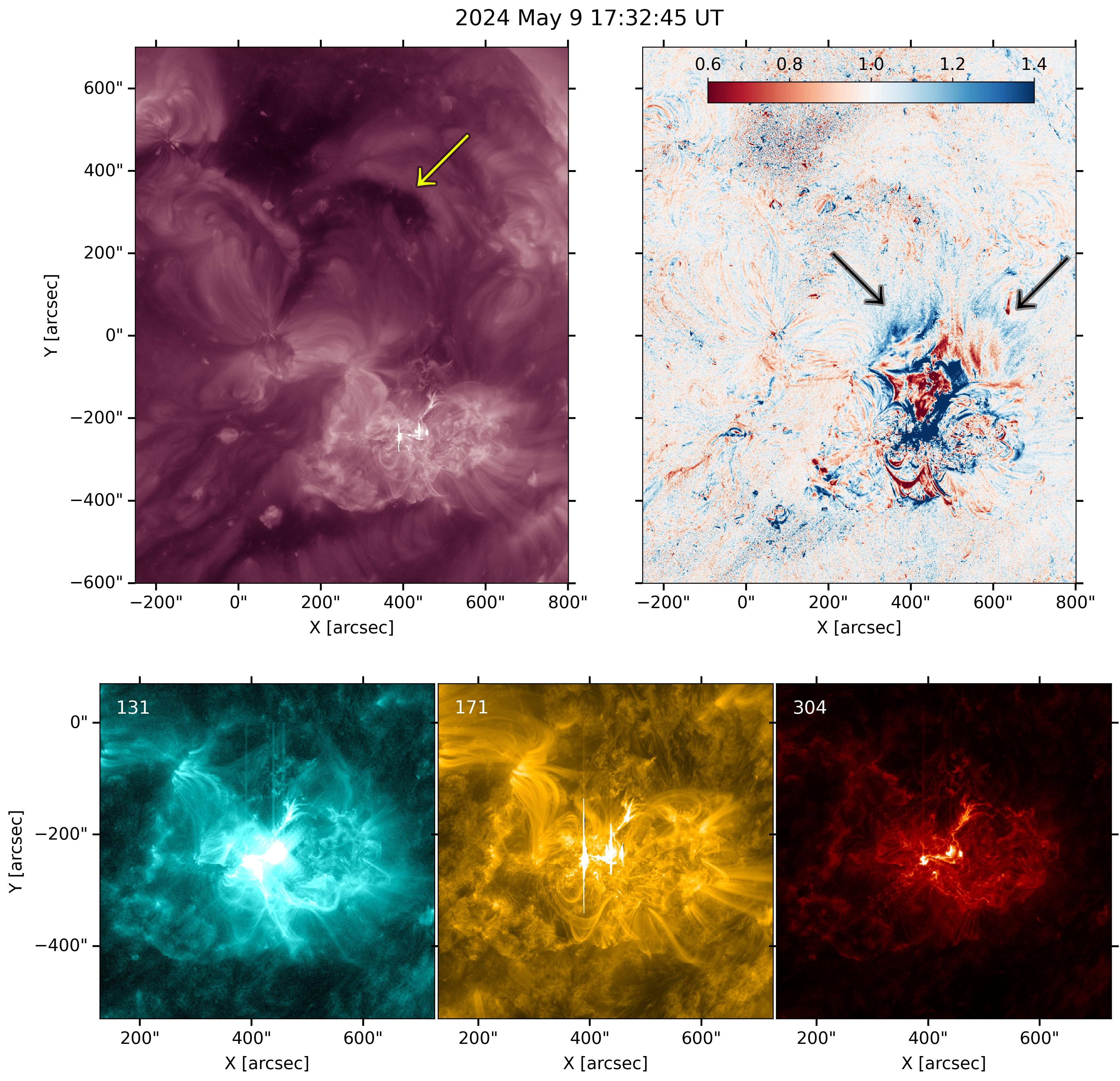}}
        \caption{Same as Fig.~\ref{fig:app:first_flare} but for the X1.1 flare on 2024 May 9 at 17:32~UT (event no.~13). The associated movie is available online.}
        \label{fig:app:second_flare}
\end{figure*}

Figures \ref{fig:app:first_flare} and \ref{fig:app:second_flare}, together with their accompanying movies, provide an observational overview of events no.~11 and 13. Each figure includes a large field of view of SDO/AIA 211~\AA~images and its base-ratio counterpart, followed by AIA 131, 171, and 304~\AA~observations focused on AR~13664. 

We note that during both events a large CH was present in the northern hemisphere and a smaller CH south of the AR, both situated to the east of AR 13664, as it can be observed in the large field AIA 211~\AA~images. The presence and location of these CHs are relevant for interpreting the large-scale magnetic connectivity and the magnetic flux systems involved in the May 2024 eruptions identified in Sect.~\ref{sec:structures}.

In event no.~11 (Fig.~\ref{fig:app:first_flare}) brightenings can be observed surrounding the D2 dimming region in the 131 and 304~\AA~observations, which are marked by yellow arrows in Fig.~\ref{fig:app:first_flare}. This is consistent with our hypothesis on the reconnection of the field lines arching from D2 to the positive polarity patch below D3 (see Fig.~\ref{fig:app:pfss_example}). The dimming region D3 is itself partially connected to the strong negative polarity sunspot northwest of the AR. Near the negative polarity sunspot, we find open flux regions (Fig.~\ref{fig:app:pfss_example}). We consider D3 to have an open-flux dimming component, as the field lines linking D3 with the northwestern region reconnect with the open field lines. In Fig.~\ref{fig:app:first_flare} brightenings and small hot loops caused by reconnection can be observed near the negative polarity sunspot in all wavelengths, and most prominently in the 131 and 304~\AA~wavelengths.

In event no.~13 an EUV wave can be observed moving northwards in the movie accompanying Fig.~\ref{fig:app:second_flare} as a blue front, corresponding to an increase in emission (marked by black arrows in Fig.~\ref{fig:app:second_flare}). As the EUV wave reaches the western arm of the northern CH (marked by a yellow arrow), reconfiguration in the quiet Sun can be observed in 211~\AA~observations. Following the passage of the EUV wave, a coronal dimming (D4 in Fig.~\ref{fig:app:pfss_example_2})  occurs in the northern hemisphere. Considering this evolution, we hypothesise that the EUV wave perturbs the local magnetic field, producing a dimming magnetically disconnected from the main eruption site (see Sect.~\ref{sec:structures}).

\twocolumn
\end{appendix}

\end{document}